\documentclass{osa-article}  
\journal{boe}

\usepackage{graphicx}
\usepackage{setspace}
\usepackage{tocloft}
\usepackage{soul}
\usepackage{siunitx}

\begin{document} 

\title{Endoscopic \textit{en-face} optical coherence tomography and fluorescence imaging using correlation-based probe tracking}

\author{Manuel J. Marques,\authormark{1,*,$\dag$} Michael R. Hughes,\authormark{1,$\dag$} Adrián F. Uceda,\authormark{1} Grigory Gelikonov,\authormark{2} Adrian Bradu,\authormark{1} and Adrian Podoleanu\authormark{1}}

\address{\authormark{1}Applied Optics Group, School of Physical Sciences, Division of Natural Sciences, University of Kent, Canterbury CT2 7NH, United Kingdom\\
\authormark{2}Institute of Applied Physics RAS, Nizhny Novgorod, Russia.\\
\authormark{$\dag$}Both authors contributed equally to this publication.} 

\email{\authormark{*}M.J.Marques@kent.ac.uk}


\begin{abstract}
    Forward-viewing endoscopic optical coherence tomography (OCT) provides 3D imaging \textit{in vivo}, and can be combined with widefield fluorescence imaging by use of a double-clad fiber. However, it is technically challenging to build a high-performance miniaturized 2D scanning system with a large  field-of-view. In this paper we demonstrate how a 1D scanning probe, which produces cross-sectional OCT images (B-scans) and 1D fluorescence T-scans, can be transformed into a 2D scanning probe by manual scanning along the second axis. OCT volumes are assembled from the B-scans using speckle decorrelation measurements to estimate the out-of-plane motion along the manual scan direction. Motion within the plane of the B-scans is corrected using image registration by normalized cross correlation. \textit{En-face} OCT slices and fluorescence images, corrected for probe motion in 3D, can be displayed in real-time during the scan. For a B-scan frame rate of 250~Hz, and an OCT lateral resolution of approximately $\SI{20}{\micro\meter}$, the approach can handle out-of-plane motion at speeds of up to 4~mm/s. 
\end{abstract}

\section{Introduction}

Endoscopic optical coherence tomography (OCT) allows  high-resolution imaging of tissue to a depth of 1-2~mm beneath the surface. It was first demonstrated as early as a few years after the first report of OCT for the eye \cite{tearney_scanning_1996}, and has since been commercialized for use in coronary arteries, complementing intravascular ultrasound for interventional guidance \cite{gora_endoscopic_2017}. There are other potential applications in the area of imaging the epithelium of internal organs, particularly for early diagnosis of cancerous lesions or for lateral margin identification during endoscopic surgery.

Endoscopic OCT probes can be broadly divided into `side-viewing' and `forward-viewing' designs. Side-viewing probes involve deflecting the beam exiting a fiber by 90 degrees, using a prism or mirror or by angle-cleaving the fiber. Scanning can be achieved by rotating the fiber using a motor at the proximal end, outside of the patient \cite{tearney_scanning_1996,rollins_real-time_1999}. Alternatively, the deflecting element may be fixed onto a micro-motor at the distal end of the probe, in which case the fiber and lens assembly itself does not need to rotate \cite{tran_vivo_2004,tsai_ultrahigh_2013,wang_heartbeat_2015}. 3D imaging is achieved by moving the probe axially, obtaining `tunnel-like' images. Side-viewing probes can have very small diameters;  needle probes with diameters as small as a few hundred microns having been demonstrated \cite{li_imaging_2000}.

Side-viewing probes are ideal for imaging tubular-shaped structures such as vasculature or parts of the gastro-intestinal tract. They are less well-suited to general endoscopic use where it may be desirable to have a forward-looking image, such as at branches of the airways, in organs such as the bladder or the stomach, or for image-guided surgery. For this reason, forward-viewing probes have also been developed, although not successfully commercialized. There are two common approaches to miniaturization of forward-viewing scanning probes. The first, and most common, is to scan the fiber itself laterally, usually behind a lens \cite{liu_rapid-scanning_2004}. It is also possible to integrate a miniature fiber-based lens into the fiber itself \cite{schulz-hildebrandt_high-speed_2018}. The second is to incorporate a miniature beam-scanning element using a micro-electromechanical (MEMS) mirror surface \cite{jung_three-dimensional_2006}, although the need to fold the beam makes this less suitable for narrow-diameter probes. In either case, the scanning can either be in one direction only, producing 2D cross-sectional OCT images, or in two directions, allowing 2D \textit{en-face} images and volumes to be assembled.

The design of forward-viewing OCT endoscopic probes is challenging in several respects. The probe would normally be introduced to the body via the working channel of an endoscope, limiting both the maximum possible probe diameter (to around 3~mm) and the length of the rigid tip. It is technically difficult to build a fast 2D scanning element at low cost and in a sufficiently small package. The field-of-view tends to be small since, unlike side-viewing probes which can rotate a full $360^\circ$, forward-viewing probes only scan over a small arc. For example, while recent state-of-the-art forward-scanning compact probe heads can acquire several volumes per second, typical image sizes are 0.8~mm \cite{schulz-hildebrandt_high-speed_2018}, 0.9~mm \cite{liang_ultrahigh_2015}, 1.3~mm \cite{wurster_endoscopic_2019}, 1.5~mm \cite{wu_miniaturized_2020}, 2~mm \cite{huo_forward-viewing_2010} and 3.5~mm \cite{sun_neurosurgical_2012}. To maintain an acceptable depth-of-field for endoscopic use, lateral resolution is typically 10-30~\textmu m, resulting in a relatively small number of lateral resolution elements in the image. 

In this paper we attempt to address some of these difficulties by introducing a new concept whereby scanning along only one axis is performed using a distal scanning mechanism, and the raster scanning (or scanning along the orthogonal axis) is completed by a manual bending mechanism of an endoscope. This relaxes the requirements on the distal scanning mechanism, and means that OCT volumes can be much larger along the manual-scan (slow-scan) direction than those that could be produced by any feasible miniaturized distal scanning mechanism within the probe itself. This new approach comes with its own challenges, particularly the difficulty of assembling artifact-free volumes during manual scanning. In our previous work using a 1D scanning endoscope and a robotic actuator to perform the slow-scan \cite{marques2019face}, volumes were assembled simply by assuming constant speed in the slow-scan direction. When using a manually-scanned endoscope, a constant scanning speed or direction cannot be assumed, and so the instantaneous velocity of the probe must be determined.

We propose here a solution involving a speckle correlation-based algorithm for determining motion in the direction out-of-plane of the B-scans, and demonstrate the feasibility of the approach in simulated imaging experiments. The scanning probe incorporates a double-clad fiber for simultaneous fluorescence imaging, and we show that we can also assemble 2D fluorescence images from a combination of mechanical and manual scanning, using the data from the OCT channel for registration. While dual fluorescence/OCT imaging has been demonstrated repeatedly using side-viewing probes \cite{yoo_intra-arterial_2011,mavadia_all-fiber-optic_2012,lorenser_dual-modality_2013,pahlevaninezhad2014high} there have been comparatively fewer forward-viewing dual-mode probes \cite{xi_integrated_2012}. Using manual scanning on one axis, we are able to obtain fluorescence images several millimeters in length.

The problem of assembling volumes from 2D B-scans has previously been studied in the context of manually-scanned 2D ultrasound probes, and algorithms based on speckle correlation have been suggested \cite{tuthill1998automated}. Conversely, most reports on the assembly of OCT images with some elements of manual scanning focus on building B-scans from  A-scans. This is in contrast to the situation where we have one axis of mechanical scanning, producing high-quality B-scans, and we wish to assemble volumes and \textit{en-face} images via manual scanning in the second, slow-scan, direction. Nevertheless, there are some clear similarities between the two tasks.

For assembly of B-scans from A-scans, in 2009 Ahmad \textit{et al.} \cite{ahmad_cross-correlation-based_2009} demonstrated a simple algorithm in which the Pearson cross-correlation coefficient is calculated between a reference A-scan and each subsequent A-scan. Once the correlation drops below an experimentally determined threshold, the current A-scan is added to the B-scan and the process is repeated (with the current A-scan becoming the new reference). To reduce the impact of structural features, the A-scan had a moving average filter of several resolution elements in size applied, and this filtered A-scan was subtracted from the unfiltered A-scan. Liu \textit{et al.} extended this work in 2012 \cite{liu_distortion-free_2012} by incorporating a theoretical model of speckle to determine the probe displacement corresponding to a specific correlation value without experimental calibration. In 2015, Wang \textit{et al.} \cite{wang_robust_2015} showed that improved performance could be achieved by taking multiple correlations between pairs of A-scans with different time separations. 

Other methods of tracking freehand-scanned probes have been suggested. For example, Ren \textit{et al.} \cite{ren_manual-scanning_2009} demonstrated 3D probe localisation to better than 20~\textmu m using a tracking camera. It is also possible to combine image-based tracking with other methods of probe localization. For example, for an OCT needle probe, Iftimia \textit{et al.} \cite{iftimia_hand_2014} combined an optical encoder for positioning with a simple algorithm to reject correlated A-scans (i.e. A-scans covering the same area of tissue) based on comparison of total intensity along the A-scan. In 2012, Yeo \textit{et al.} \cite{yeo_enabling_2012} proposed a combination of a magnetic tracking system with a correlation-based algorithm, achieving a relative positioning accuracy of 18~\textmu m. However, methods requiring additional equipment or high-quality camera images of the probe during scanning are generally undesirable if the aim is to enable simple integration with existing clinical workflows.

While these A-scan assembly techniques were developed mainly for manually-scanned OCT probes, similar techniques have been applied to proximally-driven side-viewing endoscopes in order to correct non-uniform rotational distortions (NURDs). Of particular relevance to this work,  Uribe-Patarroyo \textit{et al.} \cite{uribe-patarroyo_rotational_2015} noted difficulties with patient motion, occasional under-sampling and areas of tissue with low signal which inevitably mean that image-based tracking cannot provide a perfect correction. Aboeui \textit{et al.} \cite{abouei_correction_2018} adopted a more complex algorithm involving dynamic time warping. Other approaches to correct NURDs in side-viewing endoscopes include tracking reflections from the endoscopic sheath \cite{sun_vivo_2012} or fiducial markers \cite{ahsen_correction_2014}; these methods are not applicable to a forward-viewing probe.

The task of assembling volumes from B-scans, where the slow-scan direction is manually scanned, is also related to the task of assembling volumes in side-viewing endoscopic OCT, where  a motorized pull-back of the probe acts as the second axis of scanning. Variations in the speed of the pullback cause distortions in the resulting volume, although we expect distortions from a manually-scanned endoscope to be much larger. Lee \textit{et al.} \cite{lee_dual-beam_2017} proposed an approach in which two B-scans are acquired at two axial positions of the probe. As the probe moves axially, at some point the rear B-scan images the same point on the tissue as the front B-scan, and the time difference between these two events allows for the axial speed of the probe to be determined. A similar method was applied to a forward looking galvanometer-based \cite{harlow_dual-beam_2018} bench-top system but would be difficult to implement in a compact forward-viewing probe. More recently, Nguyen \textit{et al.} \cite{nguyen_correction_2021} proposed a method to correct both for NURDs and longitudinal speed variations in side-viewing OCT. The longitudinal correction involved analyzing the statistical variation of intensity in the \textit{en-face} image within a sliding window to estimate the pullback speed. 

As discussed below, the method we adopted is closely related to the cross-correlation techniques previously used to assemble B-scans from A-scans with manual scanning, but with the difference that mechanically-scanned B-scans are assembled into volumes (and fluorescence T-scans are assembled into fluorescence images). The lower frame rate of B-scans compared with the line rate of A-scans necessitates some modifications to the approach to allow for difficulty in ensuring over-sampling during manual scanning on the slow axis. As demonstrated below, this allows volumes to be successfully built when the probe is manually scanned using an endoscope at speeds of up to 4~mm/s.

\section{Materials and Methods}

\begin{figure}[h!]
    \centering
    \includegraphics[width=\columnwidth]{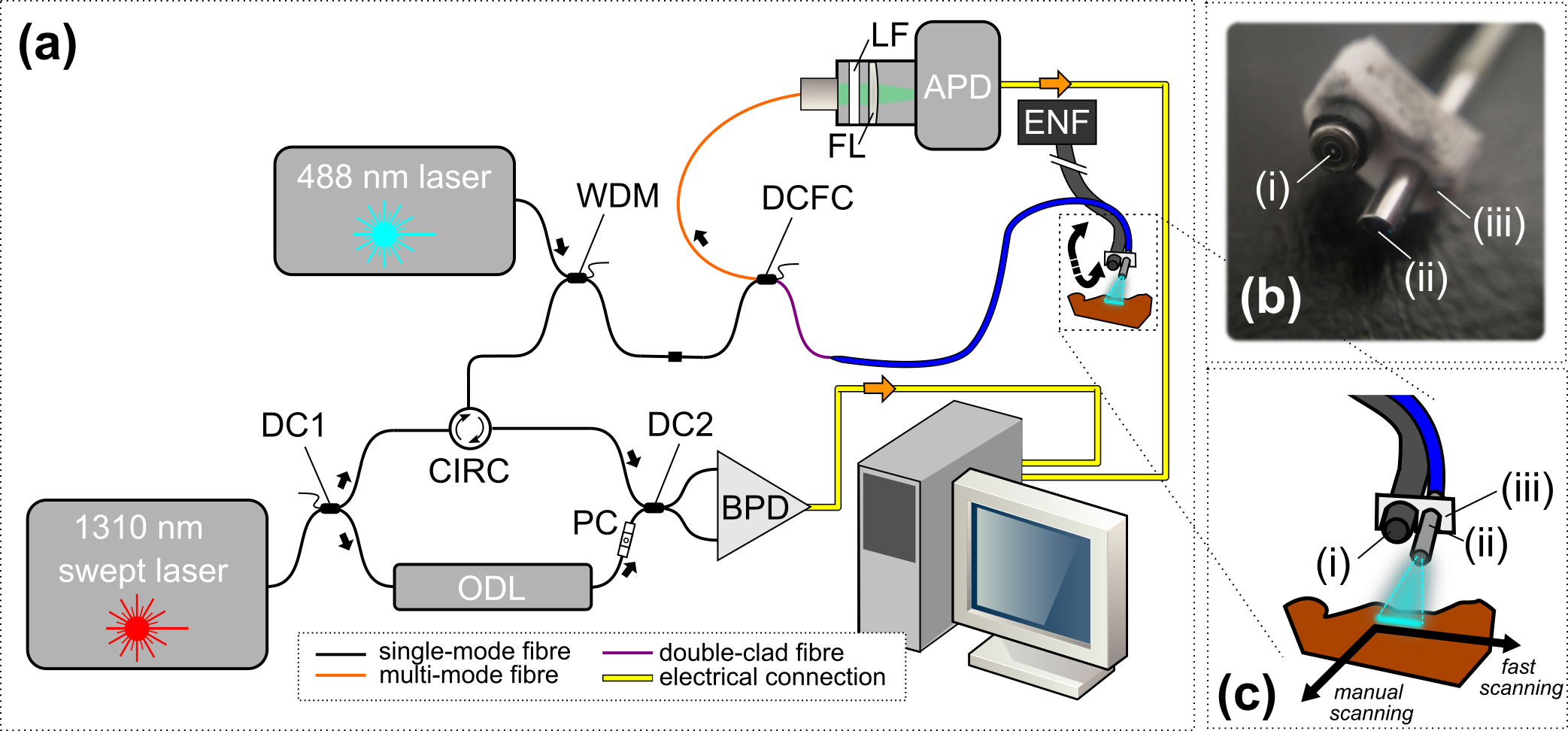}
    \caption{\textbf{(a)} Schematic diagram of the endoscopic SS-OCT/fluorescence system used in this study. \textbf{DC1-2}: fused fiber directional coupler; \textbf{CIRC}: optical fiber circulator; \textbf{WDM}: wavelength division multiplexer/combiner; \textbf{DCFC}: double-clad fiber directional coupler; \textbf{LF}: emission filter; \textbf{FL}: achromatic lens; \textbf{APD}: avalanche photo-detector; \textbf{ODL}: optical delay line; \textbf{BPD}: balanced photo-detector; \textbf{ENF}: Olympus ENF endoscope proximal controls. \textbf{(b)} Photograph of the Olympus ENF-P4 endoscope distal end (i), showing the OCT/fluorescence probe (ii) mounted to it using a 3-D printed bracket (iii). \textbf{(c)} Detail from (a), showing a close-up of the combined probe end with the Olympus endoscope.}
    \label{fig:system-diagram}
\end{figure}

A schematic diagram of the OCT/fluorescence endoscopic system used in this study is presented in Fig. \ref{fig:system-diagram}. The system, loosely based on a configuration reported by Scolaro \textit{et al} \cite{scolaro_molecular_2015}, combines a swept-source OCT sub-system operating at a central wavelength $\lambda_0 = \SI{1310}{\nano\metre}$ with a fluorescence imaging sub-system, with a $\lambda_e = \SI{488}{\nano\metre}$ solid-state laser providing the excitation. The two sub-systems share the same endoscopic probe (detailed in the next section), combined by means of a double-clad fiber coupler (DCFC, Thorlabs, model DC1300LEFA).

Briefly, in the OCT sub-system, the output from a MEMS swept-source (Axsun Technologies, central wavelength $\lambda_0 = \SI{1310}{\nano\metre}$, tuning range $\Delta\lambda=\SI{100}{\nano\metre}$, sweep rate $\SI{100}{\kilo\hertz}$) is sent to a fused fiber coupler DC1, with a splitting ratio of 90/10. 10\% of the optical power is routed to a custom-built optical delay line (ODL) forming the reference arm of the interferometer, and is afterwards reunited with the power returning from the object arm at DC2, which has a 50/50 split ratio to ensure balanced detection at the photo-detector BPD (Thorlabs, model PDB480C-AC).

In the object arm of the OCT interferometer, the optical power is routed to a fiber-based wavelength division multiplexer (WDM, Font Canada, custom $488/\SI{1310}{\nano\meter}$ fiber optic combiner) via a circulator, CIRC (AFW Technologies, model CIR-3-13-B-1-2-VR01), where it is combined with the output power from the $\SI{488}{\nano\metre}$ solid-state laser providing the fluorescence excitation (JDSU, FCD488FC-020, $\SI{488}{\nano\meter}$ wavelength, $\approx\SI{20}{\milli\watt}$ CW output power). Both are directed to the double-clad fiber coupler, DCFC (Thorlabs, model DC1300LEFA), where the combined OCT/fluorescence excitation power is routed to the double-clad fiber output, connected to the OCT/fluorescence probe.

The first cladding mode of the double-clad fiber collects any fluorescence signal from the sample, whereas the back-scattered OCT signal is routed through the core (fundamental) mode. In this way, the DCFC is capable of separating the returned OCT signal from the fluorescence signal. The latter is sent to an avalanche photo-diode amplifier module  (APD, Hamamatsu model C5460-01, frequency response from DC to $\SI{100}{\kilo\hertz}$) via an emission filter LF (Thorlabs, model FBH520-40) and an achromatic $f=\SI{200}{\milli\meter}$ lens FL (to ensure an optimum spot size on the active area of the APD), whereas the former is sent to the OCT interferometer via the WDM and CIRC.

\subsection{OCT/Fluorescence Probe}

The endoscopic probe is a forward-viewing, 1-D scanning probe, with an outer diameter of approximately \SI{3}{\milli\meter} and a working distance of approximately \SI{1}{\milli\metre}. Its working principle has been described elsewhere \cite{feldchtein2001design,cernat_dual_2012,marques2019face}; briefly, the probe operates on the voice coil principle, employing a cantilevered optical fiber which oscillates upon application of an alternating electric current. The fiber tip is imaged onto the sample by means of a gradient-index (GRIN) lens (Edmund Optics, \#64-545, $\SI{0.46}{\milli\meter}$ working distance, not shown), creating a linear scan of up to $\SI{2}{\milli\metre}$ in length, depending on the amplitude of the drive signal. Whereas the similar probe employed in our previous work \cite{marques2019face} used a single-mode, SMF-28e fiber, in this case, due to the need to detect the fluorescence signal, we instead used a 1250-1600~nm double-clad fiber (DCF13, Thorlabs, Newton, NJ, USA, core diameter $\SI{9}{\micro\meter}$, first cladding diameter $\SI{105}{\micro\meter}$, second cladding diameter $\SI{125}{\micro\meter}$).

\subsection{OCT Image Reconstruction}

In order to reconstruct OCT B-scans, the raw channeled spectra are processed using the Complex Master-Slave (CMS) method \cite{rivet_complex_2016}. CMS replaces the Fourier transform usually employed in conventional OCT signal processing, presenting advantages such as tolerance to dispersion in the interferometer \cite{bradu_demonstration_2015}, long depth imaging due to the requirement for a swept-source \textit{k}-clock being dropped \cite{marques_complex_2018}, and improved reconstruction speed when only a small number of depth points are needed \cite{bradu_recovering_2018}. This last point is of potential relevance to this work, as the image reconstruction could be made to operate over a subset of the axial range, allowing \textit{en-face} images to be assembled in real-time even for very high B-scan rates. However, this selective depth reconstruction was not necessary for the frame rates reported here. Further processing was also required to correct the distortion in the lateral direction due to the sinusoidal probe scanning drive signal. This was corrected by linear interpolation, assuming a sinusoidal scan, and was performed in real-time.

\subsection{Endoscope Integration}
\label{endoscope-int}
While the OCT probe is small enough to fit through a standard endoscope working channel, a suitable endoscope was not available when conducting this study. Instead, the OCT probe was fixed externally to an Olympus MODEL ENF-P4 endoscope, using a 3-D printed bracket as shown in Fig. \ref{fig:system-diagram}(b)(iii). The Olympus endoscope has an outer diameter of $\SI{3.6}{\milli\meter}$ and a bending range of up to 130 degrees, which allows the OCT/fluorescence probe to be scanned in the out-of-plane direction with respect to the B-scans. A similar approach of external fixing was previously employed for combined endoscopic and OCT imaging of the larynx \cite{cernat_dual_2012}. In practice the OCT/fluorescence probe would normally be inserted through the working channel of a compatible endoscope, but otherwise the procedure would be similar.

When fixing the probe to the endoscope it was necessary to ensure that the direction of endoscope bending  was approximately perpendicular to the OCT probe scanning direction. This was achieved by viewing the laser scanning line on an infra-red viewing card, rotating the OCT probe until the  line was correctly oriented, and then fixing the probe in place.

\subsection{Assembly of volumes and \textit{en-face} images - offline algorithms}
\label{sec:assembly_method}
To be able to assemble volumes or \textit{en-face} images from the raw B-scans while the probe is being manually scanned, it is necessary to estimate the instantaneous 3D velocity vector of the probe. Given that the probe is to be operated endoscopically, the exact path cannot be controlled or predicted (although the general direction of its movement can). Motion within the plane of the B-scans (axial or lateral) is straightforward to detect via image registration, while out of plane-motion is identified by looking for decorrelations in image speckle  which cannot be explained by in-plane motion. Note that is out-of-plane motion which is required by the very process of scanning - and would be deliberately introduced by the operator - in order to acquire data to assemble volumes and \textit{en-face} images. Both lateral and axial in-plane motion is undesirable but still likely to occur in practice.

A prototype of the algorithm for initial offline use was implemented in MATLAB (Mathworks). All code and a selection of example data is available on Figshare \cite{MatlabData}. A summary of the image processing algorithm is given below:

\begin{enumerate}
    \item Beginning with the first B-scan ($i = 1$), the $i$th B-scan is selected as a reference image.
    \item The average tissue surface height for the reference B-scan is identified. Several rapid methods for achieving this were investigated and are compared with manual segmentation in Supplementary Fig. S1. The method employed was binarization by thresholding of the image, followed by morphological opening and closing operations with a disk-shaped structuring element. The top surface position for each A-scan is then taken to be the location of the first non-zero pixel along the A-scan. The median of these positions across each A-scan is taken to be the average surface height for this B-scan.
    \item For the $(i+1)$th (current) B-scan image, a region of interest (ROI) of  width $\Delta x_{ROI}$ and height $\Delta y_{ROI}$ starting from a distance $d$ above the median surface height is extracted. The 2D normalized cross-correlation (NCC) is computed between this ROI and the reference image.
    \item The location of the NCC peak is identified, and this is taken as the lateral and axial shift between the two images. The current image is translated to correct for this shift.
    \item To determine the out-of-plane motion, ROIs from the reference image and the current image are high-pass filtered by subtraction of the same image with a 2D mean filter applied, similar to the method proposed by Ahmad \textit{et al.}\cite{ahmad_cross-correlation-based_2009}. This removes most structural content in the image, leaving behind the speckle pattern. A Gaussian filter is then applied to reduce noise. The ROI from the reference image is then thresholded to identify specularities and a binary mask created by applying a morphological open operation with a disc-shaped structuring element. A correlation is then performed between the pixels of the filtered ROI from the current and reference images which are not masked. This gives the degree of speckle decorrelation due to out-of-plane motion.
    \item If the correlation is not below a predefined threshold, $c_t$, indicating sufficient decorrelation for this to be identified as an independent B-scan, Steps 2-5 are repeated for subsequent B-scan images, (\textit{i} + 1), (\textit{i} + 2) ..., using the same \textit{i}th reference image until the decorrelation value drops below the threshold.
    \item If the cross-correlation peak value is below the threshold, this image is selected as the next B-scan to be used in the volume/\textit{en-face} assembly. The image is laterally and axially shifted to correct for the shifts detected in Step 4 prior to insertion into the volume.
\end{enumerate}

For the results presented below, the threshold for surface finding was set at twice the mean value for the B-scan, and the disc structuring element had a diameter of 30 pixels ($285 \times 300$~\textmu m). The high-pass mean filter had a kernel size of $5 \times 5$ pixels ($47.5 \times 50$~\textmu m), the Gaussian filter had a sigma of $1.5 \times 1.5$ pixels ($14.25 \times 15$~\textmu m), the mask threshold was 5 times the mean pixel value in the ROI, and the size of the mask structuring element was 5 x 5 pixels ($47.5 \times 50$~\textmu m). The ROI was $180 \times 100$ pixels ($1710 \times 1000$~\textmu m), laterally centered and beginning immediately below the median detected surface height. Examples of intermediate processing steps for the surface finding are shown in Supplementary Fig. S2, and for the image processing prior to the correlation calculation in Supplementary Fig. S3. The effect of the location and size of the ROI is explored in Supplementary Fig. S4.

\subsection{Handling under-sampling in the slow-scan direction by interpolation}
\label{sec:interp_method}
The above method is expected to work when the manual scanning speed is relatively slow, such that there is over-sampling in the slow-scan direction. (This is similar to the situation that occurs when manually assembling A-scans in most previous reports\cite{ahmad_cross-correlation-based_2009}). However, when the probe moves more quickly, and the degree of over-sampling is reduced, the shift will be under-estimated. This is because, when a B-scan below the correlation threshold is obtained, no account is taken of how far below the threshold the correlation has fallen. Higher speeds are therefore under-estimated and the parts of the volume/\textit{en-face} image acquired when the probe was moving faster will appear compressed in the out-of-plane direction.  A modification to the algorithm was therefore developed to calculate the location of each B-scan relative to the previously inserted B-scan based on the actual value of the correlation of the first frame to drop below the threshold. Assuming an approximately linearly fall-off in cross-correlation peak with out-of-plane distance (see Section \ref{assemblyCharacter} for analysis of this approximation), the estimated distance moved, $\Delta z$, is given by
\begin{equation}
    \Delta z = z_{dec}\bigg (1 + \frac{c_{t} - c_m}{1-c_t} \bigg)
\end{equation}

\noindent where $z_{dec}$ is the experimentally determined out-of-plane movement distance required for the B-scan correlation to drop below the decorrelation threshold, $c_{t}$, and $c_{m}$ is the measured correlation peak for this B-scan. Rather than simply inserting the B-scan as the next frame of the volume, the volume is then assembled by linear interpolation between the B-scans onto an evenly spaced grid in the out-of-plane direction. The interpolation points can be chosen to be equal to the lateral pixel size, resulting in images with square pixels, and hence the correct aspect ratio, without need for further scaling. The benefits of this approach are illustrated in Section~\ref{assemblyCharacter}.

\subsection{Surface flattening}

One of the aims of this study was to demonstrate that \textit{en-face} images can be generated in real-time by manual scanning along the slow axis. \textit{En-face} images are generally more useful if they follow the shape of the tissue surface rather than being an oblique slice through the tissue. As described above, axial shifts between B-scans are already corrected by the algorithm. While this was designed primarily to correct for probe axial shifts during acquisition, it will also tend to smooth out  undulations in the tissue surface in the slow-scan direction, since these two effects are not easily separated. In addition to this, we also explored the possibility of correcting for a constant probe tilt along the fast scan axis, since again, if this is not corrected, this tilt would result in the \textit{en-face} image being an oblique slice through the tissue. This is done by correcting for the average tilt of the tissue surface across all the B-scans of the volume. A volume was first assembled without lateral shift corrections. The volume was then re-sliced to obtain a stack of B-scans oriented along the slow scan direction. The median surface heights along each of these re-sliced B-scans (i.e. for each position in the fast scan) are then determined using the same method as described above. A linear fit to these surface heights gives an estimate of the probe tilt in the fast-scan direction. This measured tilt is then used to provide a simple correction for probe tilt (by axially shifting each A-scan of each B-scan) as part of a full reconstruction which includes both axial and lateral shifts. 

\subsection{Assembly of volumes and \textit{en-face} images - online algorithms}

To demonstrate the feasibility of assembling \textit{en-face} images in real-time during endoscopic investigations, a simplified form of the algorithm was implemented in LabVIEW (National Instruments). While the majority of the algorithm was implemented using standard LabVIEW functions (including the Vision Development Module), cross-correlation and image filtering prior to speckle correlation measurement was performed using the OpenCV4 library, accessed from LabVIEW via DLLs compiled from C++. The implementation functioned similarly to the method described above, including tissue flattening in the slow-scan direction and interpolation. Probe tilt correction (i.e. flattening in the fast-scanning lateral direction) was not performed, as this requires information from the entire volume scan and hence cannot be implemented in a real-time \textit{en-face} preview. OCT B-scans can be streamed directly to the \textit{en-face} assembly algorithm from the OCT acquisition and processing software or from a saved datafile. Images were successfully processed in real-time at B-scan frame rates of 250~Hz, and the \textit{en-face} image extracted at a fixed depth relative to the estimated tissue surface was displayed with line-by-line update as the probe was scanned. A video of the real-time display is provided in Supplementary Video 1 \cite{RTvideo} and the LabVIEW code is provided on Figshare \cite{LabviewCode}.

\section{Quantitative Characterization and Validation}

\subsection{Optical Coherence Tomography Probe and System}
\label{sec:characterisation}
The procedure to quantify the sensitivity of the OCT channel was the same as in a previous study from Marques \textit{\textit{et al.}} \cite{marques2019face}. An A-scan was obtained from a brushed metal target and the signal to noise ratio (SNR) of the A-scan peak was measured with respect to the average noise floor to be $\SI{39.8}{\decibel}$. The reference arm was then blocked and the optical power was measured at the BPD first from the brushed metal and then from a mirror. The ratio of these measured powers was $\SI{35.5}{\decibel}$, and this was used to correct the measured signal-to-noise ratio of the signal from the metal to obtain a sensitivity of $\SI{75.3}{\decibel}$. This is much lower than $85-\SI{92}{\decibel}$ in our other MS systems previously reported in the literature. This may be due to lower efficiency in re-injecting light back into the DCF inside the custom-made probe. The width of the A-scan peak was measured to be $\SI{28}{\micro\meter}$, which effectively dictates the axial resolution of the system.

\begin{figure}[h!]
    \centering
    \includegraphics[width=0.9\columnwidth]{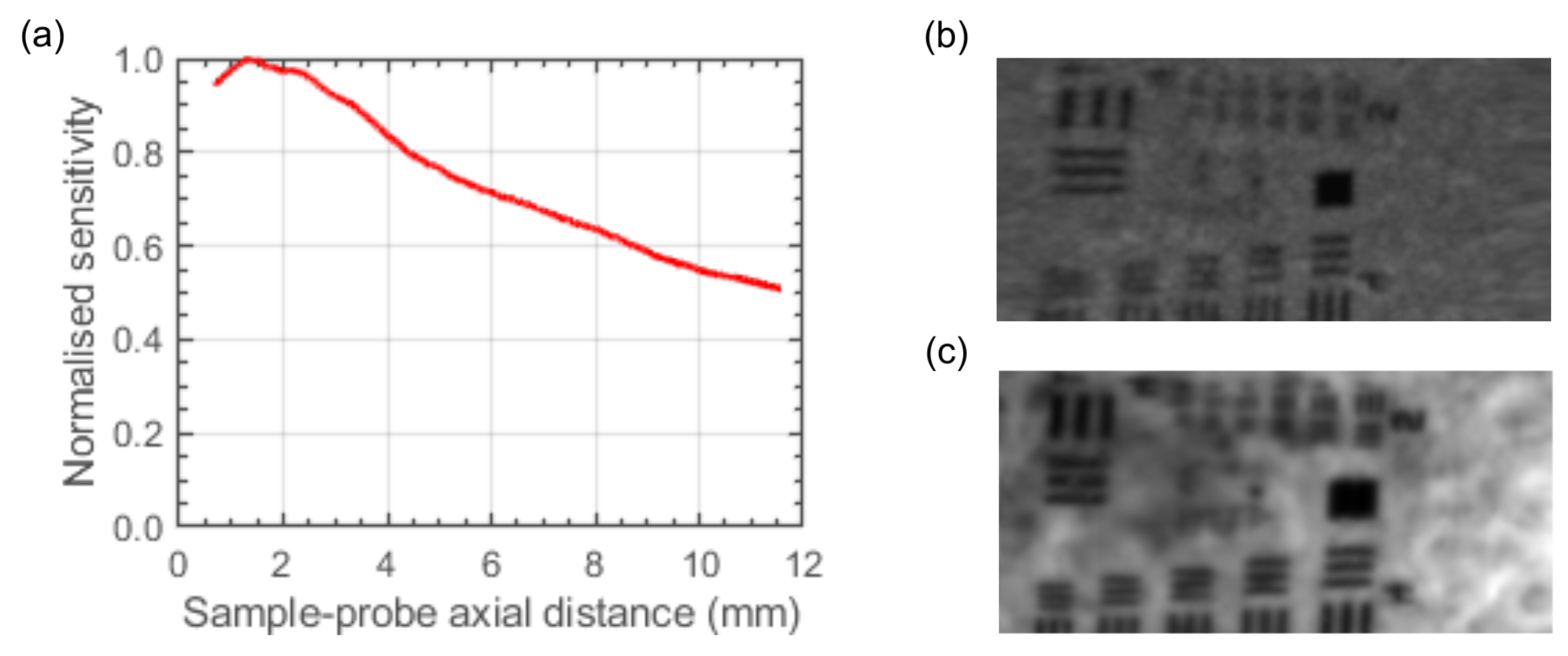}
    \caption{(a) Fluorescence signal as a function of distance from probe tip. A moving average filter of 5 points was applied to smooth the fluorescence sensitivity profile. (b) \textit{En-face} OCT image of a positive USAF target. (c) Fluorescence image of positive USAF target placed over fluorescently-stained paper.}
    \label{fig:fluo-sensitivity}
\end{figure}

To measure the signal level of the fluorescence channel as function of distance from the probe, the probe was mechanically translated at constant speed away from a piece of paper stained with a 0.05g/100ml solution of acriflavine hydrochloride. The average signal across the fluorescence T-scan is plotted as a function of distance from the probe tip (determined from the position of the OCT-scan surface peak) in Fig. \ref{fig:fluo-sensitivity}(a). Peak signal was obtained at 1.5 $\pm$ 0.1 mm from the probe, and there is no significant selection in depth for the fluorescence channel.

To characterize the lateral resolution of the probe, a positive United States Air Force (USAF) resolution target was placed on top of the stained paper and the probe was mechanically translated across the sample.  An \textit{en-face} slice from the OCT channel and a fluorescence image are shown in Fig.~\ref{fig:fluo-sensitivity}(b) and (c), respectively. In the OCT channel, element 4 of group 4 can be resolved, giving a lateral resolution of 22.1 \textmu m. In the fluorescence channel,  element 1 of group 5 can be resolved, giving a lateral resolution of 15.63 \textmu m. 

\subsection{Characterization of Volume/\textit{en-face} Assembly}
\label{assemblyCharacter}

To determine the relationship between speckle correlation and out-of-plane motion, the probe was translated over a scattering phantom (paper resolution target) and chicken breast tissue at 0.5~mm/s using a motorized translation stage, perpendicular to the fast-scan direction (i.e in a direction that is out-of-plane of the OCT B-scan). The normalized cross correlation between the first image and a region of interest of $180 \times 100$ pixels ($1710 \times 1000$~\textmu m) extracted from each subsequent image was then calculated to determine in-plane shifts. The shift-corrected images were then processed as described above in Section~\ref{sec:assembly_method}, and the correlation as a function of out-of-plane motion distance was then calculated. This was performed 50 times, each with a different starting point within the scan. The correlation for zero shift was calculated by averaging the correlation between 20 pairs of images when the probe was not in motion to account for the effect of random noise. 

\begin{figure}[h!]
    \centering
    \includegraphics[width=\columnwidth]{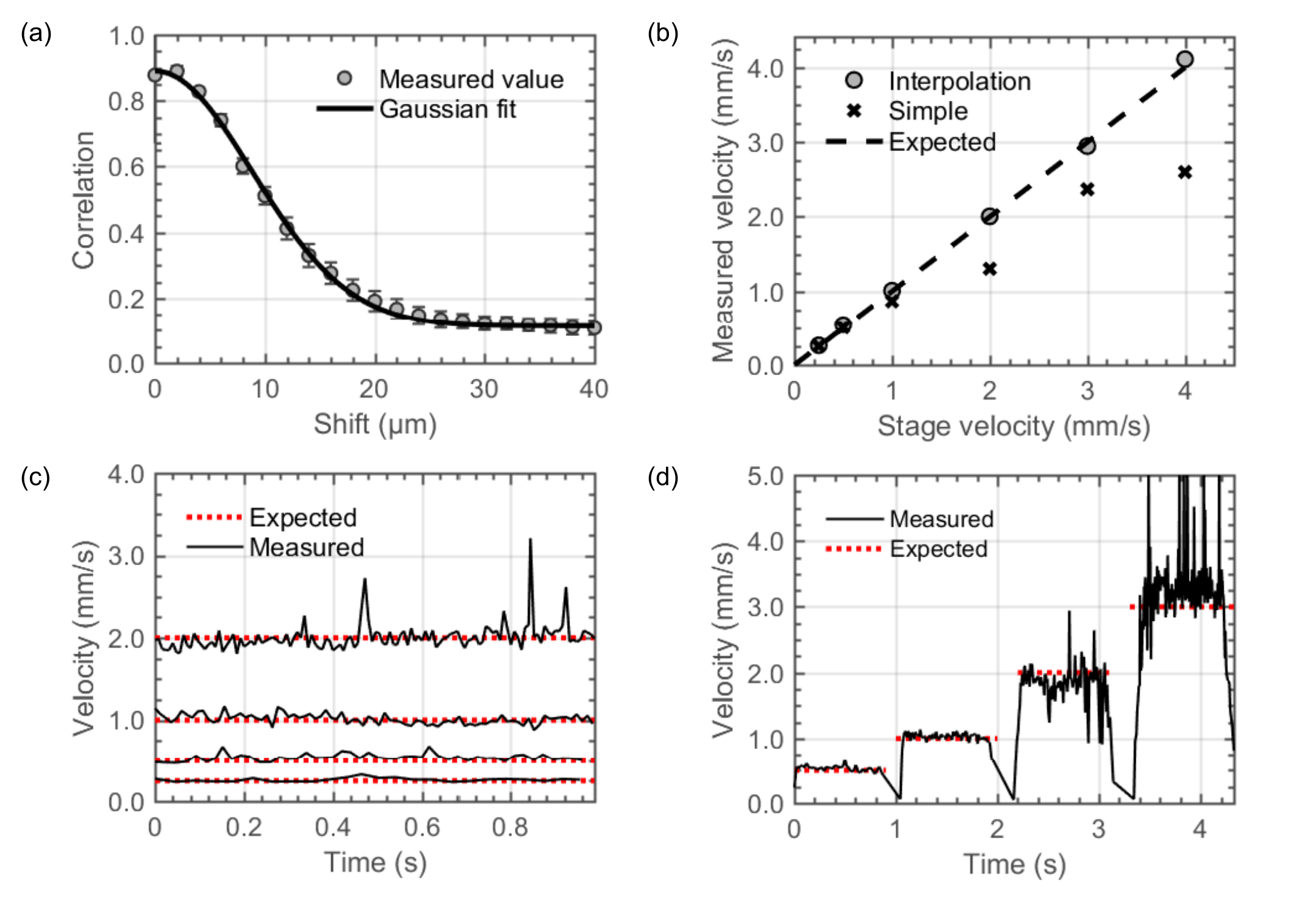}
    \caption{Validation of speed estimation in the out-of-plane direction via speckle decorrelation, using mechanical translation stage to move the OCT probe over chicken breast tissue. (a) Correlation as a function of out-of-plane movement distance, average of 50 starting points. Error bars are standard deviation across 50 runs. A least-squares  Gaussian fit is shown. (b) Measured speed using simple and interpolation based methods for six different velocities. (c) Speed measured as a function of time using interpolation method for four different translation stage speeds. (d) Speed measured as a function of time using interpolation method for probe motion with varying speed. Drops to zero speed are a mechanical feature of the way in which the stage was programmed, and are not artifacts of the method. }
    \label{fig:velocity-calib}
\end{figure}

The mean correlation is shown as a function of out-of-plane distance for the chicken breast tissue in Fig.~\ref{fig:velocity-calib}(a). The correlation drops to 0.5 at a distance of 10.4~\textmu m. The curve is approximately linear between correlations of 0.7 and 0.3, which supports the use of the linear interpolation method described in Section~\ref{sec:interp_method}. For the paper target, the correlation drops to 0.5 at a distance of 10.8~\textmu m, suggesting that this calibration is not highly dependent on the sample properties, and so this data is not shown. 

A Gaussian fit to the data of the form
\begin{equation}
    f(x) = (f_{max} - f_{min})\exp(-x/2\sigma^2) + f_{min}    
\end{equation}
is also shown. $\sigma$, $f_{max}$ and $f_{min}$ are free parameters, found to be 12.0~\textmu m, 0.952 and 0.052 for paper and 12.5~\textmu m, 0.891 and 0.114 for the tissue, respectively. $f_{max}$ is the correlation between images when there is no out-of-plane motion, $f_{min}$ is the residual correlation at large distances, and $\sigma$ is an estimate of the speckle grain size and hence the lateral resolution.  Using the conventional definition of resolution as the full-width half-maximum (FWHM) of the point-spread function, this corresponds to an expected lateral resolution of approximately 21~\textmu m in the out-of-plane direction, compared to the experimentally determined value of 22~\textmu m in the in-plane direction.

This curve can be used to determine an appropriate value for the correlation threshold, $c_t$. Clearly the value must lie in the range between $f_{min}$ and $1$ or it will never be reached. As the threshold approaches $f_{min}$ the separation between B-scans will become reasonably large compared to the lateral resolution, while a threshold close to 1 will be more subject to random noise or other variations unrelated to out-of-plane motion. To follow the convention of sampling approximately twice per resolution element, a threshold of $c_t = 0.5$ was adopted for the following work. The out-of-plane movement required to reach this decorrelation threshold is then $z_{dec} = 10.4$~\textmu m.

The speed of the probe in the out-of-plane direction can be estimated by determining how many B-scan frames are required for the correlation to drop below the threshold ($n$). The measured average speed, $\bar{v}$, is then given by
\begin{equation}
    \bar{v} = \frac{Rz_{dec}}{n}
\end{equation}
where $R$ is the B-scan frame rate. The number of frames, $n$, can be calculated using either the simple method, taking the first frame where the correlation falls below the threshold, or the interpolation method, in which a non-integer number of frames is calculated using the method described in Section~\ref{sec:interp_method}. For both methods, Fig.~\ref{fig:velocity-calib}(b) shows the average estimated speed when the stage was set to move at several constant speeds. The simple method under-estimates larger speeds quite significantly as expected. The maximum possible speed which can be predicted by the simple method is that corresponding to when a single frame ($n=1$) is sufficient to drop below the threshold, which is approximately 2.6~mm/s with the parameters used here. In comparison, the interpolation method continues to accurately estimate speeds as high as 4~mm/s, even though the probe is under-sampling by a factor of 1.5 at this speed.

The instantaneous speed estimated as a function of time is shown in Fig.~\ref{fig:velocity-calib}(c) for four velocities. The standard deviation is approximately 10\% of the mean for all velocities in the range 0.25 to 4~mm/s, but there are an increased number of spurious high and low speed points for the larger speeds. Figure~\ref{fig:velocity-calib}(d) shows the results of setting the motorized stage to change speed during the acquisition. Due to the way the stage was programmed, the speed briefly dropped to zero between each segment, as can be seen in the estimated speed. The processing step of masking bright pixels prior to measuring the correlation between images was found to be critical when specular reflections occurred from the tissue surface, otherwise the probe speed would be significantly under-estimated at these points, as shown in Supplementary Fig. S5.

\begin{figure}[h!]
    \centering
    \includegraphics[width=1\columnwidth]{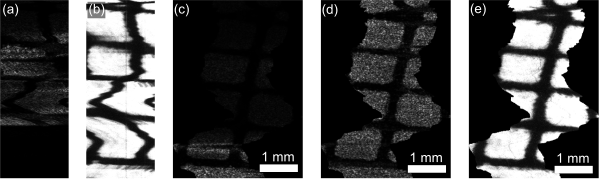}
    \caption{Reconstruction of OCT and fluorescence images following freehand probe scan over fluorescently-stained printed grid phantom. (a) \textit{En-face} slice extracted from raw volume. (b) Raw fluorescence image. (c) \textit{En-face} slice from motion-corrected volume without surface correction. (d) \textit{En-face} slice from motion-corrected volume with surface correction. (e) Motion-corrected fluorescence image. (a)-(b) have the same horizontal scale as (c)-(e) but have no vertical scale since this depends on the instantaneous probe speed.}
    \label{fig:calib-phantom}
\end{figure}

The OCT probe was then moved freehand over a printed grid phantom with fluorescent highlighter applied. An example of the raw \textit{en-face} OCT and fluorescence images are shown in Fig.~\ref{fig:calib-phantom}(a,b). Severe distortion is present due to lateral motion and inconsistent out-of-plane motion, as expected. Figure~\ref{fig:calib-phantom}(a) also shows the effect of axial motion of the probe; this is less apparent in Fig.~\ref{fig:calib-phantom}(b) due to the lack of optical sectioning in the fluorescence channel (see Section~\ref{sec:characterisation}). Figure~\ref{fig:calib-phantom}(c) shows the reconstructed \textit{en-face} image correcting only for out-of-plane and lateral motion, while Fig.~\ref{fig:calib-phantom}(d) shows the result if surface-flattening is also employed. Finally, Fig.~\ref{fig:calib-phantom}(e) shows the reconstructed fluorescence image. While some artifacts can be observed in the reconstructed \textit{en-face} views, the algorithm successfully recovers the broad morphology of the sample.

\section{Imaging Results Under Simulated Conditions}

\begin{figure}[h!]
    \centering
    \includegraphics[width=1\columnwidth]{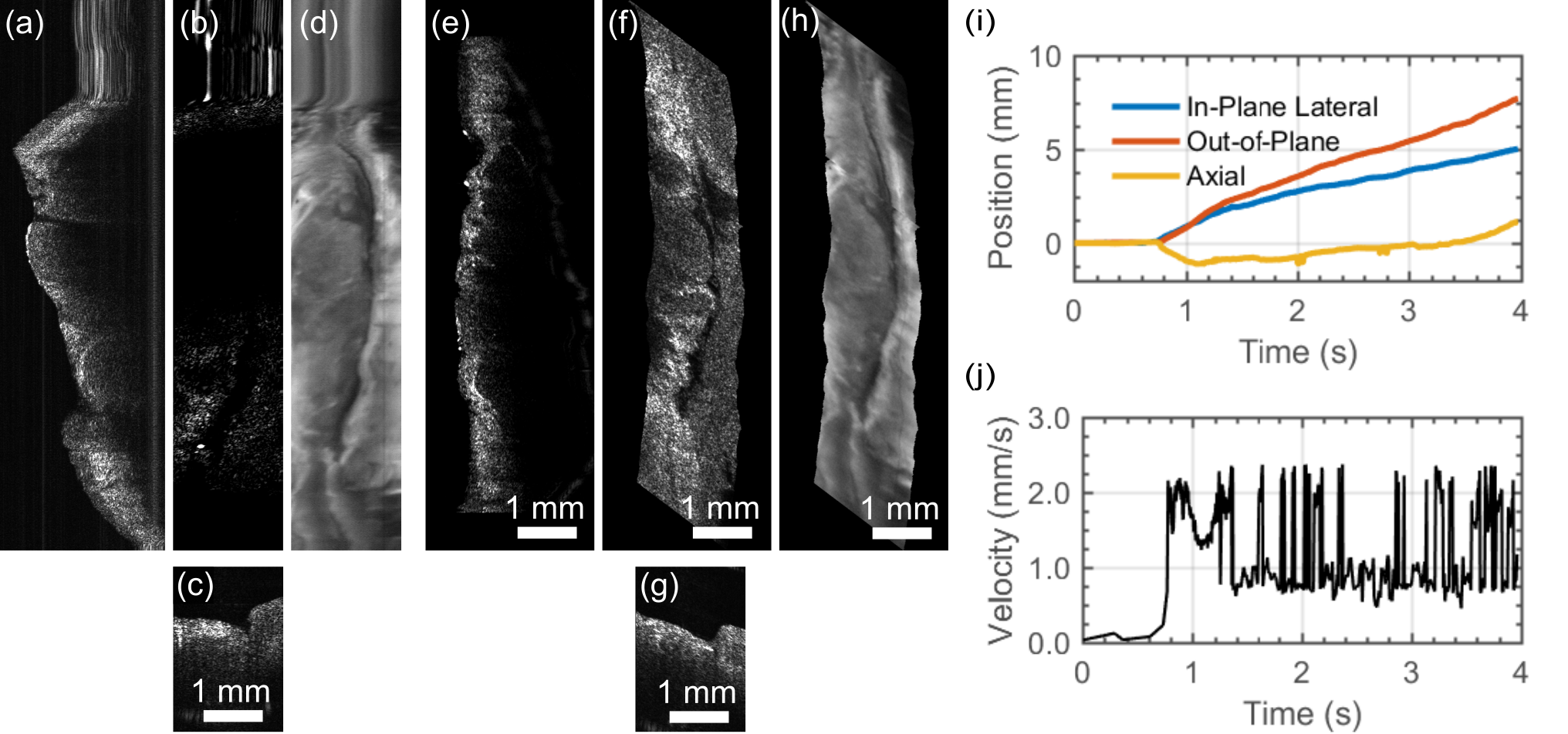}
    \caption{Example reconstruction using endoscope for slow-axis scanning over porcine lung tissue. The raw volume is represented in (a), (b) and (c), which shows a B-scan along the slow-scan direction (a), an \textit{en-face} slice (b) and a B-scan along the fast-scan direction (i.e. a raw B-scan) (c). The raw fluorescence data is also shown in (d). The reconstructed OCT volume is shown in (e), (f) and (g), which again shows a slow-axis B-scan (e), an \textit{en-face} view (f) and a fast-axis B-scan (g). The reconstructed fluorescence is shown in (h). The \textit{en-face} and fluorescence images have been rotated by $37^\textrm{o}$ for display purposes. The in and out-of-plane lateral and axial shifts detected by the algorithm are shown in (i), and the measured out-of-plane speed is shown in (j). (a) and (b) are at the same horizontal scale as the other images, but have no vertical scale since the vertical position corresponds to the time of acquisition of each B-scan.\label{fig:endoscope-analysis}}
    
\end{figure}

An example of an OCT volume and fluorescence images of porcine lung tissue, acquired using the endoscope for the slow-scan (as described in Section~\ref{endoscope-int}), is shown in Figure~\ref{fig:endoscope-analysis}. To provide fluorescence signal, the tissue was stained for approximately 2 minutes with acriflavine hydrochloride before rinsing with water. Figure~\ref{fig:endoscope-analysis}(a-d) show raw acquisitions, while Fig.~\ref{fig:endoscope-analysis}(e-g) show reconstructed images after processing. Figure~\ref{fig:endoscope-analysis}(c,g) show B-scans along the fast-scan direction, while Fig.~\ref{fig:endoscope-analysis}(a,e) show B-scans extracted from the volume along the slow-scan direction. Figure~\ref{fig:endoscope-analysis}(b,f) show \textit{en-face} images extracted from the volume. There was considerable in-plane lateral motion during this scan, and so the \textit{en-face} and fluorescence images have been rotated by 37 degrees for convenient display. The effect of the algorithm in removing the motion artifacts near the top of the images can clearly be seen. The surface flattening effect can also be clearly observed by comparing the slow-axis B-scans and by observing the more constant intensity along the \textit{en-face} image in Fig.~\ref{fig:endoscope-analysis}(f).

Figure~\ref{fig:endoscope-analysis}(i) shows the relative probe position as a function of time calculated by the algorithm, while Fig.~\ref{fig:endoscope-analysis}(j) shows the calculated speed in the slow-scan (out-of-plane) direction. For most of the scan the measured probe speed in the out-of-plane direction was approximately 1~mm/s, although there are a large number of spikes up to around 2~mm/s which could be a result of jerky motion leading to sudden decorrelation.

\begin{figure}[h!]
    \centering
    \includegraphics[width=1\columnwidth]{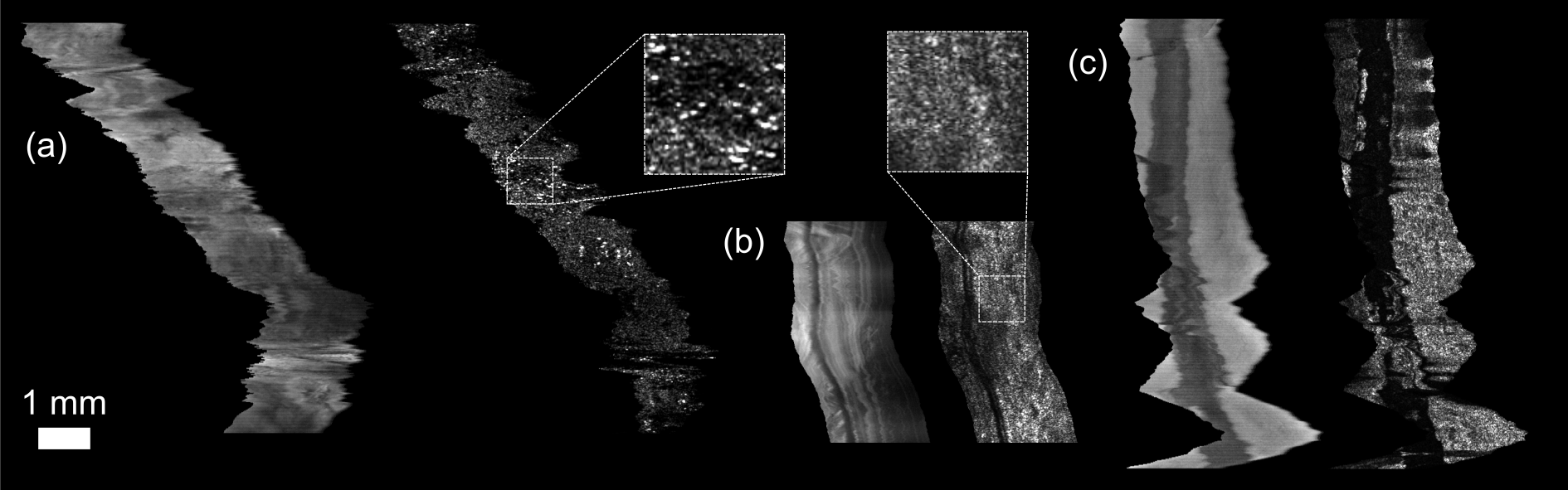}
    \caption{Representative pairs of fluorescence (left) and OCT images (right) from porcine \textit{ex vivo} tissue generated when using an endoscope for the slow-axis scanning. (a) Lung, (b) esophagus, and (c) trachea. The samples were stained with arcriflavine hydrochloride for 2 minutes and then rinsed prior to imaging. \textit{En-face} slices were manually selected for display from reconstructed volumes. For display purposes, \textit{en-face} OCT slices were contrast adjusted using the ImageJ autocontrast tool. Fluorescence images were autocontrasted to show the full dynamic range of the image. Zoomed insets show 1x1~mm regions.}
    \label{fig:endoscope-tissue}
\end{figure}

Further example of images acquired using the endoscope are shown in Fig.~\ref{fig:endoscope-tissue}. \textit{En-face} and fluorescence images are shown from porcine trachea in Fig.~\ref{fig:endoscope-tissue}(a), lung in Fig.~\ref{fig:endoscope-tissue}(b) and esophagus in Fig.~\ref{fig:endoscope-tissue}(c). The tissue was prepared and stained as before.  All three sets of images are shown at the same scale, except for the zoom insets which show an area of 1~mm~x~1~mm.

\section{Discussion and Conclusions}
We have demonstrated the feasibility of using a conventional endoscope to provide the slow-axis scan for simultaneous 3D OCT and fluorescence endoscopic imaging. In experiments designed to simulate clinical imaging, slow-axis scans over ranges of up to 1~cm were performed, and \textit{en-face} images successfully reconstructed using the speckle decorrelation and registration algorithms. While some image artifacts are often present, and the method is not robust to certain kinds of motion, the \textit{en-face} images generally appear congruent and are much larger in the slow-scan direction than could be achieved using any miniaturized 2D endoscopic scanning mechanism.

The reconstruction performance is primarily limited by the fast-axis scanning rate of the endoscopic OCT probe. With a B-scan frame rate of 250~Hz and a lateral resolution of approximately 20~\textmu m, the (manual) slow-axis scan is limited to a speed of approximately 2.5~mm/s before under-sampling. At speeds greater than 5~mm/s the adjacent B-scans will be almost entirely decorrelated and it is no longer possible to estimate the speed correctly. These scanning speeds are rather low and may be difficult to achieve in clinical practice. However, a relatively modest increase in the speed of the B-scan frame, for example to 1 kHz (which is technically feasible), would increase the permitted scanning speed to between 10~mm/s and 20~mm/s. The maximum speed is therefore not a limitation of the approach in general. Sudden or jerky motion during a scan will also lead to missed areas of tissue, and even with very high fast-axis scanning speeds it is unlikely that these artifacts could be avoided entirely. 

The manual slow-scan direction must be roughly perpendicular to the fast-scan direction. While some in-plane motion parallel to the fast-scan direction can be tolerated, and is corrected by the registration procedure, large amounts of in-plane motion will reduce the accuracy of the speckle decorrelation algorithm and lead to a smaller area of tissue being imaged. It would therefore not be possible to simply insert the OCT probe into the working channel of an endoscope; there would need to be some way to check that the orientation aligns with one axis of the endoscope bending motion. This could be intrinsic, through mechanical design forcing the probe to be inserted correctly, or by visualizing the fluorescence excitation on the endoscope camera view and manually rotating the probe. Alternatively, it may be desirable to permanently build the OCT probe into the endoscope, it which case the alignment would be fixed during manufacture.

The speckle decorrelation algorithm cannot detect a change in the direction of the out-of-plane motion. While in principle the endoscope operator could ensure that the endoscope is translated only along a single direction during the scan, patient motion or inadvertent endoscope motion could result in the same area of tissue being scanned over more than once. However, it may be possible to detect and remove these occurrences through image analysis.

These limitations are partially mitigated by allowing the operator to see the \textit{en-face} image being assembled in real-time, providing visual feedback on the scanning speed and direction and allowing any errors to be more easily identified as they occur. We have demonstrated that the algorithm is computationally inexpensive and can readily be applied in real-time for a B-scan rate of 250~Hz. To allow for higher frame rates, Complex Master-Slave OCT could be used to reconstruct only a limited axial range around the known surface height.

We have therefore shown that manual slow-axis scanning in combination with correlation-based probe tracking is a promising approach for endoscopic forward-viewing OCT and fluorescence imaging. To aid further investigation, we have made the code for reconstructing OCT volumes and fluorescence images available with this report. The approach may also have other applications, such as for lower-cost hand-held probes for external body imaging or industrial inspection and non-destructive testing. Alternatively, it may  be applied to endoscopic probes designed to be operated with mechanical or robotic scanning systems (such as robotic surgical systems). Indeed, for robotic systems it should be possible to use the speed estimation from the OCT speckle decorrelation for closed-loop robot control, leading to more precise and controlled imaging.

\begin{backmatter}
\bmsection{Funding}
To be included from Prism.


\bmsection{Disclosures}
AP and AB are co-inventors of patents US10760893 and US9383187 in the name of the University of Kent.

\bmsection{Data Availability}
Data and Matlab code underlying the results presented in this paper are available on Figshare \cite{MatlabData}. The real-time LabVIEW code is also available on Figshare \cite{LabviewCode}.

\bmsection{Supplemental document}
See Supplement 1 for supporting content. 
\end{backmatter}




\end{document}


\maketitle
\section{Comparison of surface detection methods}
The method to determine the out-of-plane speed requires that a region of interest (ROI) is extracted from within a scattering region of the sample. In practice this means placing the ROI just under the top surface. We compared five different approaches for determining the location of the top surface of the sample in the B-scan:
\begin{itemize}

\item \textbf{Method 1.} Convolution of each A-scan in the  B-scan with a step function (5 pixels of -1, 5 pixels of 1). The position of the peak intensity of the result for each A-scan is then taken as the position of the surface for that A-scan.

\item \textbf{Method 2.} As Method 1, but using correlation instead of convolution.

\item \textbf{Method 3.} As Method 2, but using normalized cross correlation. 

\item \textbf{Method 4.} Binarization of the B-scan (using a threshold of twice the mean pixel value for the image) followed by morphological open and close operations with disc-shaped structuring element of $30 \times 30$ pixels ($285 \times 300$~\textmu m). The first non-zero pixel in each A-scan is then taken to be the surface position for that A-scan. 

\item \textbf{Method 5.} Applying a Sobel filter and then taking the position of the peak intensity in the result for each A-scan as the location of the surface position for that A-scan.

\end{itemize}

For each approach, three different pre-processing steps were also tested:  an auto intensity adjustment (Matlab imadjust, using 1\% saturation high and low), a Wiener filter, and a 3x3 moving average filter. 

\begin{figure}[h!]
    \centering
    \includegraphics[width=\columnwidth]{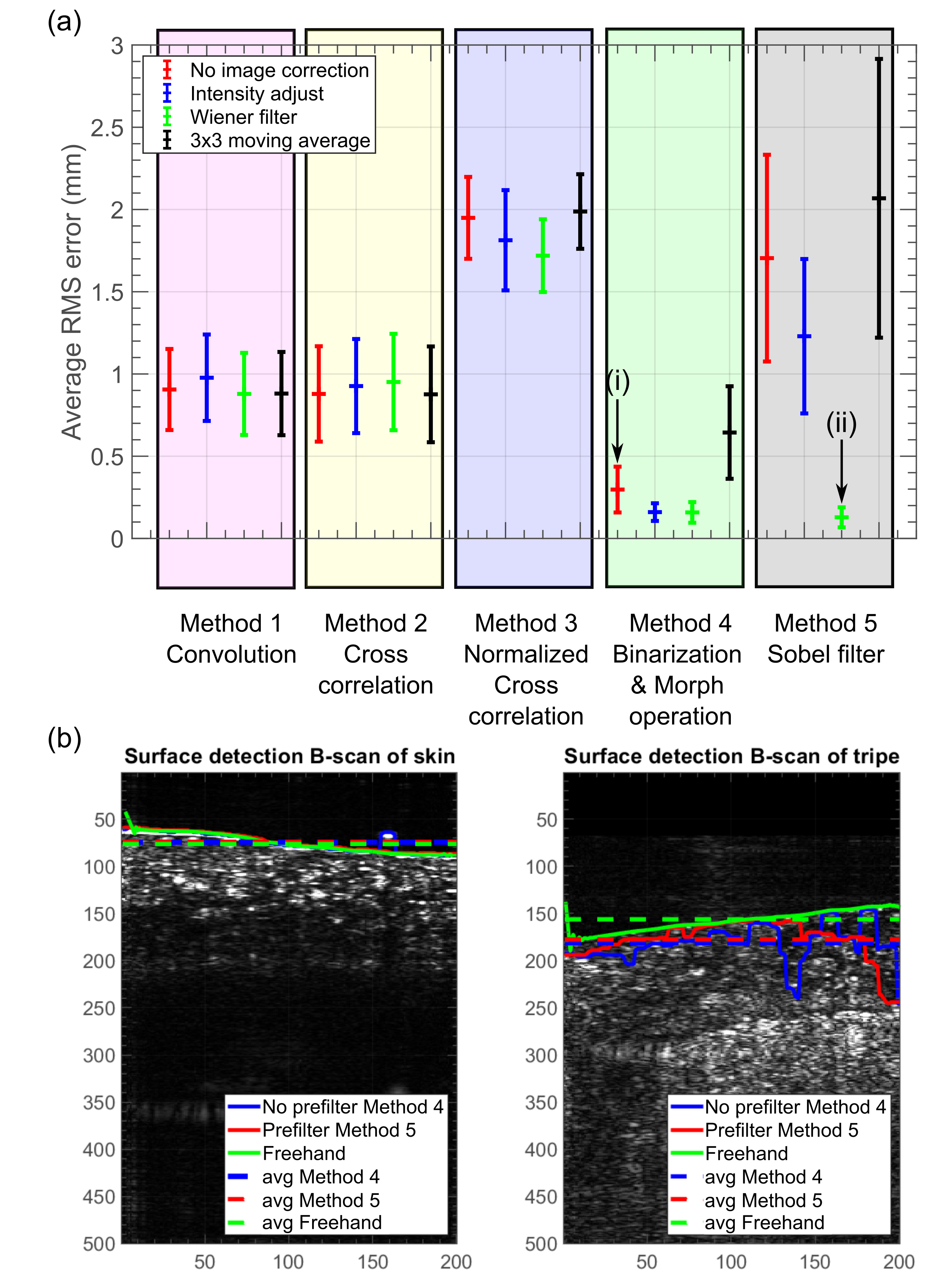}
    \caption{(a) Average RMS error with respect to manual segmentation for each combination of surface detection approach and image pre-correction. The error bars are the standard deviation of the RMS values across 30 different test images. Arrows indicate: (i) Method 4 with no pre-correction, which was employed in the paper, and (ii) Method 5 with a Wiener pre-filtering, which provides slightly better results at the expense of more complex processing. (b) Examples of B-scans of human skin (left) and ox tripe (right). The dashed lines indicate the median position of the surface across the sample while the continuous lines show the position detected at each A-scan. The freehand registration is shown in green, the Sobel method in red, and the binarization in blue. The units are image pixels, in the vertical scale 1 pixel is 10~\textmu m, in the horizontal scale 1 pixel is 9.5 \textmu m.}
    \label{fig:surface_comparison}
\end{figure}

All methods were compared against manual segmentation of a test set of B-scans extracted from 10 data volumes, with three B-scans analysed per volume. The first three volumes were taken from \textit{in-vivo} human skin imaging, while the other seven were taken from \textit{ex-vivo} ox tripe. The RMS difference between the manual and automatic surface position was then averaged across all the datatsets. The results, grouped by method and image pre-correction, are shown in Fig.~\ref{fig:surface_comparison}(a).

The Sobel filter with pre-processing using the Wiener filter (Fig.~\ref{fig:surface_comparison}(a)(ii)) resulted in the lowest average RMS error across the datasets (\hl{0.13 mm}) . However, while this was slightly better than the  binarization and morphological operations method (\hl{0.29 mm}), the binarization method is computationally much simpler. Small errors with respect to the manual segmentation are not of great importance as the ROI will still cover approximately the desired area of tissue. The binarization and morphological method (Fig.~\ref{fig:surface_comparison}(a)(i)) was therefore used throughout this study.

Fig.~\ref{fig:surface_comparison}(b) compares the results of these two methods  on the images on which they performed best and worst:  an example of \textit{in-vivo} human skin, on the left, and \textit{ex-vivo} tripe, on the right. In both cases, the median top surface position (and hence the location used for the top of the ROI) is the same to within a few pixels (i.e. a few 10s of \textmu m).

\section{Examples of intermediate processing steps}
\begin{figure}[h!]
    \centering
    \includegraphics[width=\columnwidth]{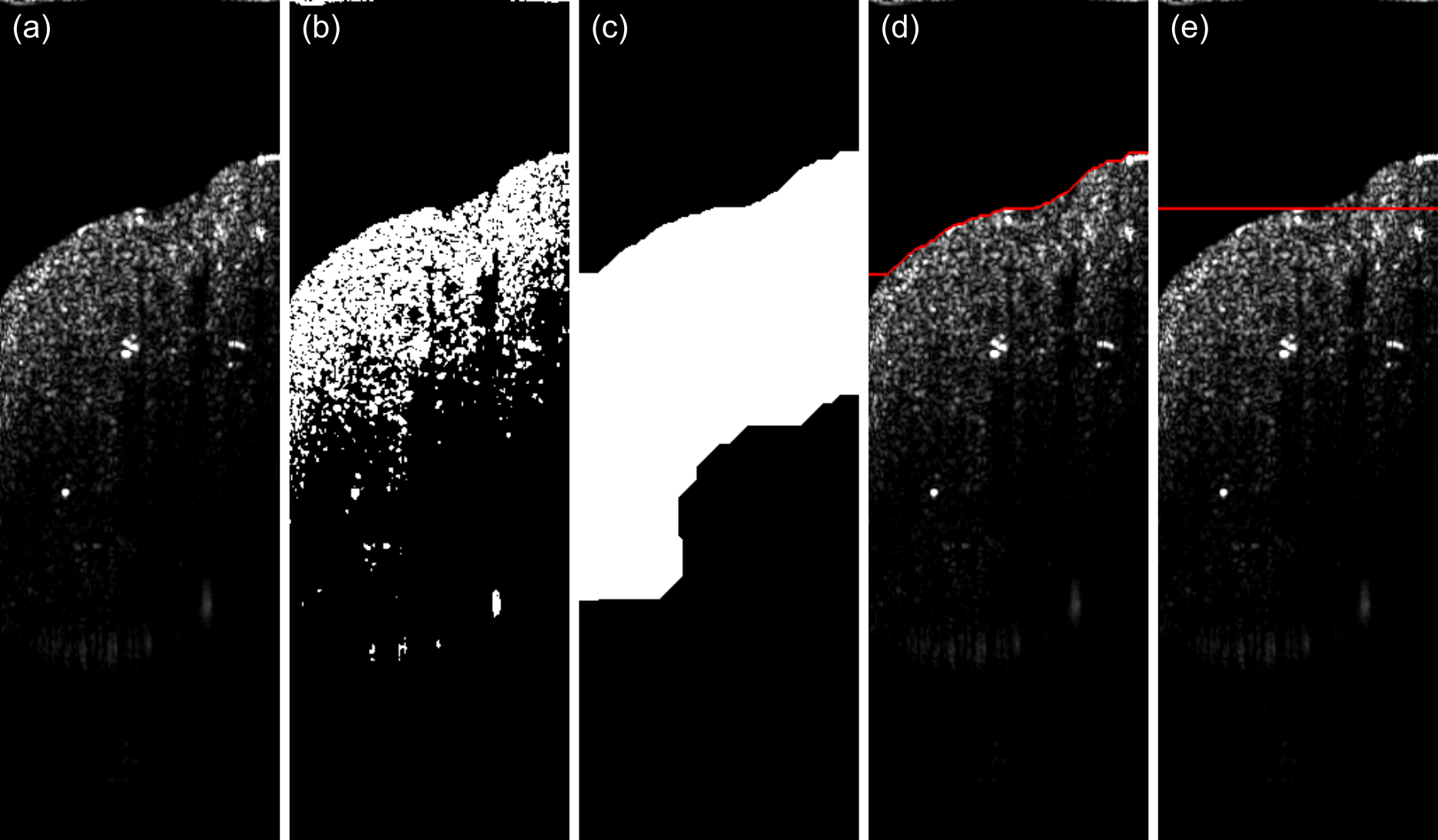}
    \caption{Example of surface finding in a B-scan of porcine lung tissue. (a) B-scan. (b) Thresholded and binarized B-scan. (c) Morphological close and open. (d) Estimated surface overlaid on B-scan (e) Median surface overlaid on B-scan.}
    \label{fig:surf-steps}
\end{figure}

As an example, the processing steps for locating the approximate axial position of the tissue surface in a B-scan using the selected method are shown for \textit{ex vivo} imaging of porcine lung tissue in Fig.~\ref{fig:surf-steps}. The raw image shown in Fig.~\ref{fig:surf-steps}(a) is thresholded and binarized, as shown in  Fig.~\ref{fig:surf-steps}(b), and morphological open and close operations are performed, as shown in Fig.~\ref{fig:surf-steps}(c). The location of the first non-zero pixel in each column is shown overlaid on the B-scan in Fig.~\ref{fig:surf-steps}(d). The median value of these surface heights, which is used to position the region of interest (ROI) for correlation measurements, is shown in Fig.~\ref{fig:surf-steps}(e).

\begin{figure}[h!]
    \centering
    \includegraphics[width=0.9\columnwidth]{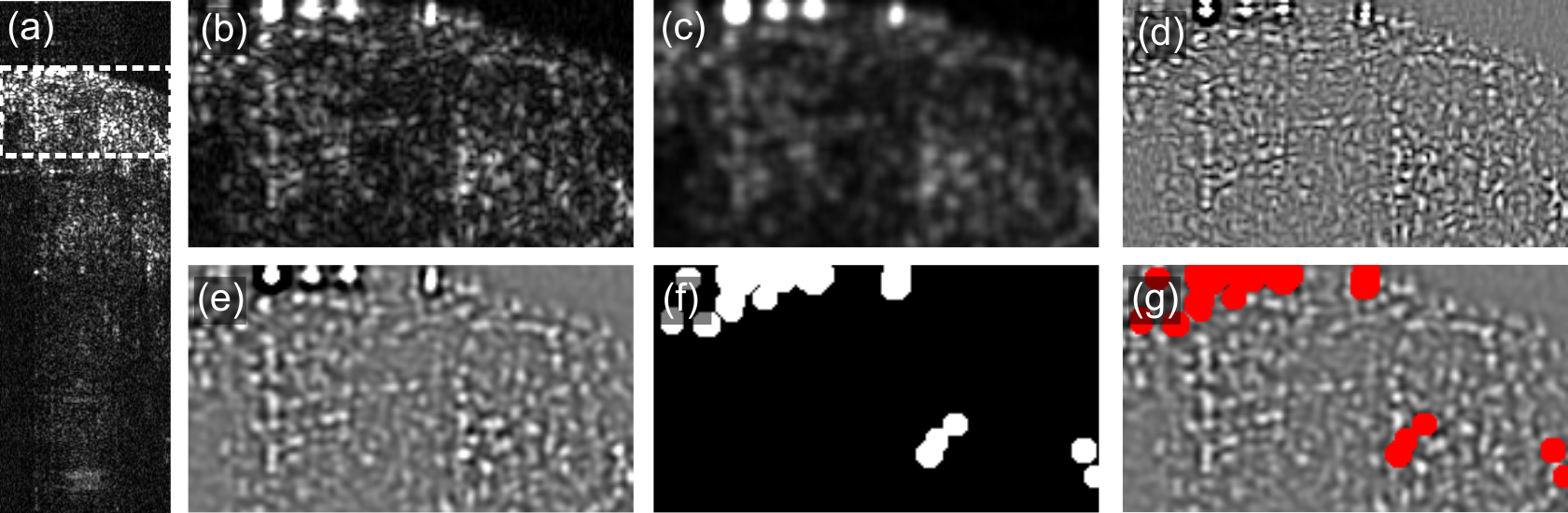}
    \caption{Example of processing steps prior to correlation calculation. (a) Complete B-scan of chicken breast tissue showing region of interest (ROI) as dashed box. (b) Extracted ROI. (c) Mean-filtered ROI. (d) Result of subtracting mean-filtered ROI from original ROI. (e) Result of a applying 2D Gaussian filter to (d). (f) Mask calculated by thresholding and morphological open operation on (b). (g) Mask overlaid on the filtered ROI.}
    \label{fig:proc_steps}
\end{figure}

Pre-processing steps prior to computing the correlation between two images are illustrated in Fig.~\ref{fig:proc_steps}. The full B-scan shown in Fig.~\ref{fig:proc_steps}(a) has a ROI extracted, as shown in Fig.~\ref{fig:proc_steps}(b). The mean filtered version of this ROI, Fig.~\ref{fig:proc_steps}(c) is subtracted from the original ROI to obtain an image with increased speckle contrast, Fig.~\ref{fig:proc_steps}(d). This is then Gaussian filtered to reduce noise, as shown in Fig.~\ref{fig:proc_steps}(e). Thresholding, followed by a morphological open operation, produces a mask which is shown in Fig.~\ref{fig:proc_steps}(f) and overlaid as red pixels on the ROI in Fig.~\ref{fig:proc_steps}(g). Only pixels which are not masked, i.e. those which are black in Fig.~\ref{fig:proc_steps}(f), are used in the correlation calculation, to avoid saturated pixels affecting the results

\section{Effect of ROI position}
\begin{figure}[h!]
    \centering
    \includegraphics[width=\columnwidth]{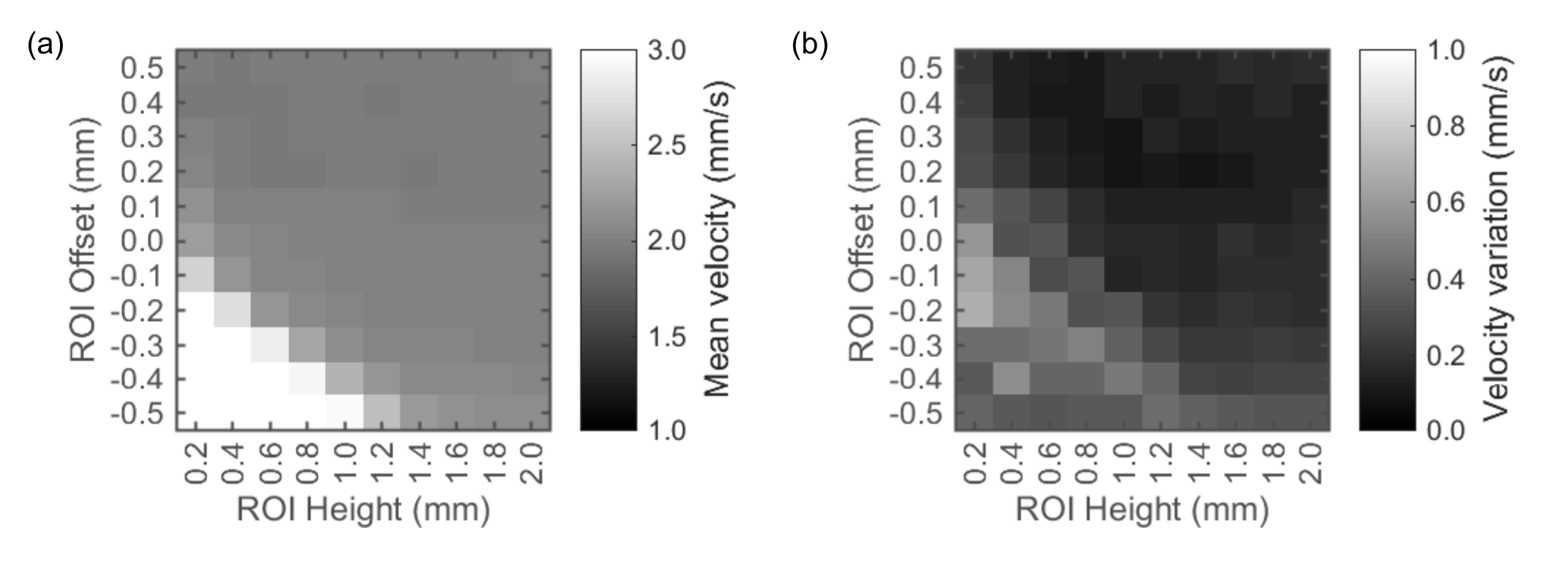}
    \caption{Effect of vertical (i.e. axial) position and size of region of interest (ROI) on measured out-of-plane speed. Negative offsets indicate a ROI higher in the image. The probe was translated at 2~mm/s for 1~s over chicken breast tissue using a translation stage. (a) Mean of measured speed. (b) Standard deviation of measured speed. }
    \label{fig:roi_position}
\end{figure}

The boundaries of the region of interest (ROI) for correlation between two images are selected based on the average surface height for the first of those images. Since the surface topography will vary, it is important that the correlation and hence the estimate of the probe shift is not overly sensitive to the exact axial position of the ROI. Figure~\ref{fig:roi_position} shows the effect of adjusting the vertical position and size of the ROI for a dataset collected with the probe moved by a translation stage for 1~s at a speed of 2~mm/s over \textit{ex-vivo} chicken breast tissue. The measured speed, shown in Fig.~\ref{fig:roi_position}(a) is approximately 2~mm/s everywhere except in the lower left corner of the plot which represents ROIs which lie entirely or mainly above the surface of the tissue. The standard deviation of the measured speed over the course of the scan is similar for all combinations which do not result in ROIs largely above the surface. For results reported in the paper, a ROI height of 1~mm and offset of 0~mm was used.

\section{Effect of masking specularities}
\begin{figure}[h!]
    \centering
    \includegraphics[width=\columnwidth]{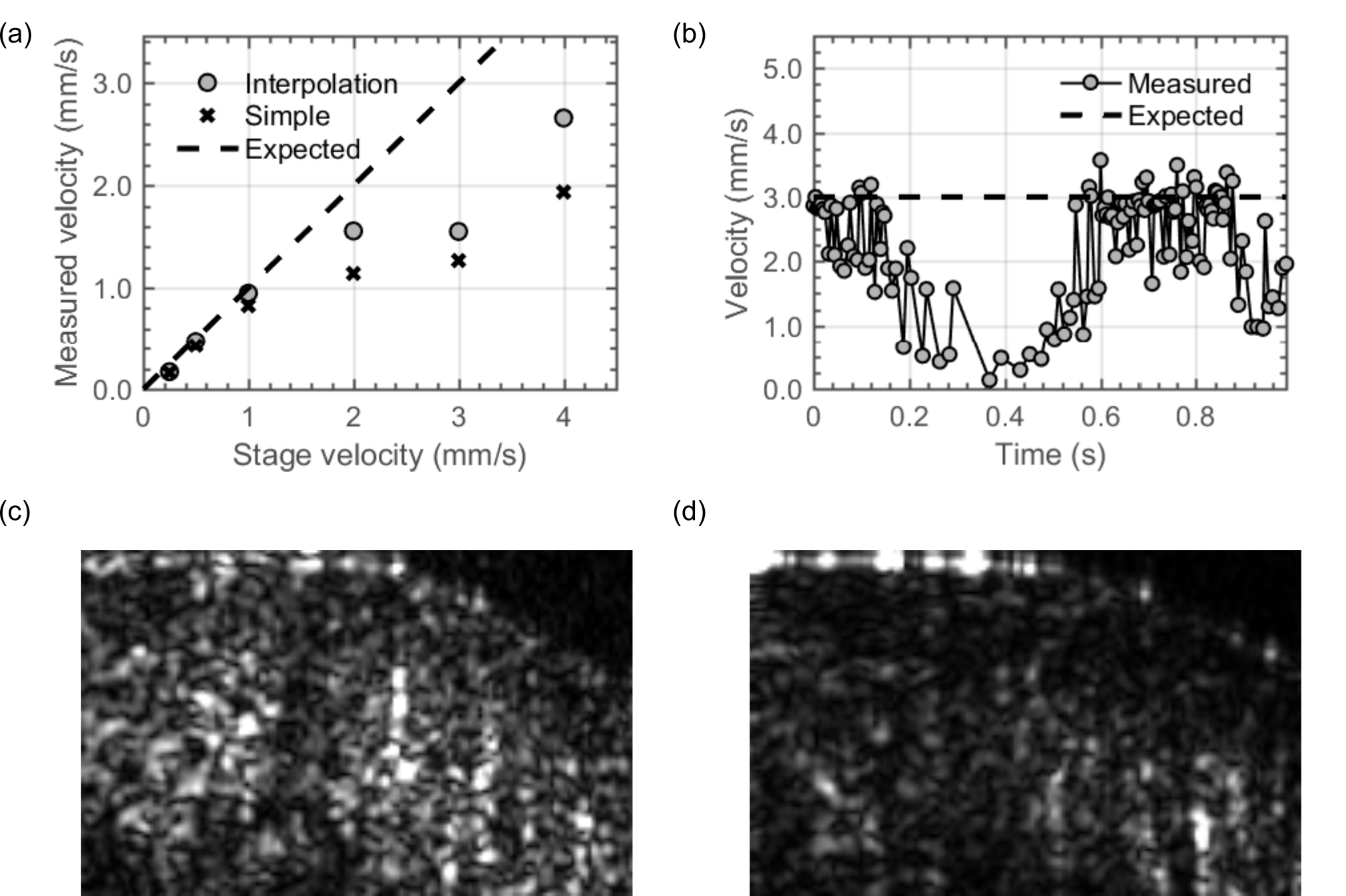}
    \caption{Out-of-plane speed measured when masking is not used. Probe was translated at several different speeds using a motorized translation stage over chicken breast tissue. (a) Measured speeds using both simple and interpolation-based methods for six different translation stage speeds. (b) Speed measured as a function of time using the interpolation method for a translation stage speed of 3~mm/s. (c) ROI from the B-scan image acquired at time 0.88~s. (d) ROI from the B-scan image acquired at time 0.48~s. The images in (c) and (d) were both independently contrast-adjusted to show their full dynamic range.}
    \label{fig:masking}
\end{figure}

The procedure to remove bright pixels from the correlation measurement is essential when the surface presents specular reflections. This is illustrated in Fig.~\ref{fig:masking}, which shows the results of processing the same speed calibration data as used for Fig.~3 but in this case without masking of bright pixels prior to calculating correlations. From Fig.~\ref{fig:masking}(a) it can be seen that the average measured out-of-plane speeds for several of the programmed stage speeds were grossly under-estimated. Figure~\ref{fig:masking}(b) shows the measured speed as a function of time when the stage was moved at 3~mm/s, where it can be seen that the measured speed is highly variable (mean of 1.7~mm/s, standard deviation of 0.35~mm/s). This can be understood by inspection of the ROIs extracted from B-scans acquired at times of 0.88~s in Fig.~\ref{fig:masking}(c) and at 0.48~s in Fig.~~\ref{fig:masking}(d). For the frame at 0.48~s, there are obvious specular reflections at the tissue surface which dominate the correlation measurement and result in a slower decorrelation with distance (and hence lower measured speed). When masking is performed, these bright pixels on the surface are not included in, and hence do not affect, the correlation calculation. Note that the two images have been independently contrast-adjusted for display purposes.

